\documentclass[preprint]{aastex}

\usepackage{graphicx,amsmath}

\usepackage{caption}
\usepackage{xcolor}
\usepackage{graphicx}
\usepackage{subfigure}
\usepackage[final]{pdfpages}
\usepackage{url}
\usepackage{natbib}
\usepackage{color}
\usepackage{pdflscape}
\usepackage{hyperref}
\usepackage{float}
\usepackage{soul}
\setstcolor{blue}
\usepackage{tikz}
\usepackage{ amssymb }
\usetikzlibrary{shapes,arrows}
\let\new=\newcommand 
\new{\diff}{{\rm d}}
\new{\be}{\begin{equation}}
\new{\ee}{\end{equation}}

\slugcomment{Journal of Astrophysics and Astronomy (Accepted)}

\shorttitle{The $M_{\bullet} - \sigma$ relation in spherical systems}
\shortauthors{Bhattacharyya \& Mangalam}

\begin{document}

\title{The $M_{\bullet} - \sigma$ relation in spherical systems}

\author{Dipanweeta~Bhattacharyya$^{\dagger}$,~~~~A.~Mangalam$^{\ddagger}$}

\affil{Indian Institute of Astrophysics, Bangalore-560034, India}

\email{ $^{\dagger}$dipanweeta@iiap.res.in, $^{\ddagger}$mangalam@iiap.res.in}

\begin{abstract}
To investigate the $M_\bullet -\sigma$ relation, we consider realistic elliptical galaxy  profiles that are taken to follow a single power law density profile given by $\rho(r) = \rho_{0}(r/ r_{0})^{-\gamma}$ or the Nuker intensity profile. We calculate the density using Abel's formula in the latter case by employing the derived stellar potential in both cases, we derive the distribution function $f(E)$ of the stars in presence of the supermassive black hole (SMBH) at the center and hence compute the line of sight (LOS) velocity dispersion as a function of radius. For the typical range of values for masses of SMBH, we obtain  $M_{\bullet} \propto \sigma^{p}$ for different profiles. An analytical relation $p = (2\gamma + 6)/(2 + \gamma)$ is found which is in reasonable agreement with observations (for $\gamma$ = 0.75 - 1.4, $p$ = 3.6 - 5.3). Assuming that a proportionality relation holds between the black hole mass and bulge mass, $ M_{\bullet} =f M_b$, and applying this to several galaxies we find the individual best fit values of $p$ as a function of $f$; also by minimizing $\chi^{2}$, we find the best fit global $p$ and $f$. For Nuker profiles we find that $p$ = $3.81 \pm 0.004$ and $f$ = $(1.23 \pm 0.09)\times 10^{-3}$ which are consistent with the observed ranges.
\end{abstract}


\section{Introduction}

It is now widely accepted that all massive galaxies have supermassive black holes at their centers. At distances close to the centers of these galaxies stellar or gas motions are completely dominated by the gravity of the SMBH than that of the nearby stars. The relation of the SMBHs to their host galaxies can be seen by the strong correlation between the mass of SMBH and velocity dispersion $\sigma$ of the stars in the galaxy. This is somewhat surprising because the stars are too far from the SMBH for the velocity dispersion to be affected by its gravitational field. Its origin is still a topic of debate. 

The $M_{\bullet}- \sigma$ relation is given by the equation
\begin{eqnarray}
M_\bullet = k \sigma^p, \label{m_sigma_rel}
\end{eqnarray}
which was first reported by \cite{2000ApJ...539L...9F} with the index $p$ = $4.8 \pm 0.5$, whereas, \cite{2000ApJ...539L..13G} reported $p$ = 3.75 $\pm$ 0.3. The former used symmetric linear regression method for their analysis and in this process both the variables $M_{\bullet}$ and $\sigma$ had a unique error in measurements as well as intrinsic scatter, while the later used non - symmetrical least square regression, where it was assumed that $\sigma$ had no uncertainty in measurement and $M_{\bullet}$ had same uncertainty for all. This relation is observed in ellipticals and evolved bulges. \cite{2013ApJ...765...23D} claim that their latest measurements indicate that there is no evidence of offset for this relation between ellipticals and classical bulges. Table \ref{msigma} shows the values of the indices determined by different authors using different techniques.

\begin{table}[H]
\begin{center}
\scalebox{0.75}{
\begin{tabular}{|c |c |c |c |}\hline
\# & References & $p$  &  Comments\\ \hline\hline
1 &\cite{2000ApJ...539L...9F} & 4.8 $\pm$ 0.5 & 12 elliptical galaxies with known $\sigma$ \\ \hline
2 & \cite{2000ApJ...539L..13G} & 3.75 $\pm$ 0.3 & 26 galaxies with measured $M_{\bullet}$ and $\sigma$\\ \hline
3 & \cite{2001ApJ...547..140M} & 4.72 $\pm$ 0.36 & 27 galaxies with measured $M_{\bullet}$ and $\sigma$\\ \hline
4 & \cite{2002ApJ...578...90F} & 4.58 $\pm$ 0.52 & 16 spirals and 20 elliptical galaxies\\ \hline
5 & \cite{2002ApJ...574..740T} & 4.02 $\pm$ 0.32 & 31 galaxies with measured $M_{\bullet}$ and $\sigma$\\ \hline
6 & \cite{2005SSRv..116..523F} & 4.86 $\pm$ 0.43 & SMBHs which have resolved $r_{h}$\\ \hline
7 & \cite{2009ApJ...698..198G} & 4.24 $\pm$ 0.41 & Combination of spiral and elliptical galaxies \\ \hline
8 & {\em ibid} & 3.96 $\pm$ 0.42 & 25 Elliptical galaxies \\ \hline
9 & \cite{2013ARAA..51..511K} & 4.38 $\pm$ 0.29 & Classical bulges and ellipticals \\ \hline
10 & \cite{2013ApJ...764..184M} & 5.64 $\pm$ 0.32 & 19 late type and 53 early - type galaxies\\ \hline
11 & \cite{2013ApJ...764..151G} & 5.53 $\pm$ 0.34 & 51 non - barred galaxies\\ \hline
12 & \cite{2013ApJ...765...23D} & 4.39 $\pm$ 0.42 & Sample of \cite{2009ApJ...698..198G} with newly measured $M_{\bullet}$ and $\sigma$ \\ \hline
13 & \cite{2017ApJ...838L..10B} & 4.76 $\pm$ 0.60 & 32 quiescent galaxies \\ \hline
14 & {\em ibid} & 3.90 $\pm$ 0.93 & 16 AGN host galaxies \\ \hline
\hline \hline
\end{tabular}
}
\caption{A survey of the $M_{\bullet}-\sigma$ relation giving the historical determinations of the slopes.}
\label{msigma}
\end{center}
\end{table}

There are many theoretical models proposed for explaining the $M_{\bullet} - \sigma$ relation. \cite{1998A&A...331L...1S} discussed an energy driven flow where the energy released in accretion is used completely in the process of unbinding the mass of the galactic bulge. They modeled a protogalaxy and assumed it to be an isothermal sphere to find $p=5$. \cite{2003ApJ...596L..27K} has described another model based on gas accretion but with momentum driven flow, giving $p=4$. In this process, it has been assumed that cooling by inverse Compton occurs and thereby a fraction of energy is lost to radiation while some part of the accretion energy is available for unbinding the bulge. At a point, the gas accretion stops and the black hole mass saturates; this is the maximum mass attained by gas accretion in presence of cooling. \cite{2002NewA....7..385Z} based on the loss cone dynamics in an isothermal halo obtained $p=5$ for a model based on growth by stellar ingestion. The origin of this relation is still a mystery but various models give a range $p=4 - 5$, which is in rough agreement with observations.


From observations, it is seen that bulge mass, $M_{\bullet} \simeq f M_{b}$, where, $f$ = 1.259 $\times 10^{-3}$ \citep{2001ApJ...547..140M}. Later \cite{2003ApJ...589L..21M} and \cite{2004ApJ...604L..89H} found $f=2\times 10^{-3}$ and $f=(1.4\pm0.4)\times 10^{-3}$ respectively. For higher masses, the relation is said to be nonlinear and given by $M_{\bullet}\propto M_{b}^{1.12}$ \citep{2004ApJ...604L..89H} where the Jeans equation has been applied with zero anisotropy in the system to determine the velocity dispersion of 30 elliptical galaxies whose bulge masses are sourced from \cite{1998AJ....115.2285M}. \cite{2013ARAA..51..511K} have also found the following relation:
\begin{eqnarray}
\bigg(\frac{M_{\bullet}}{10^{9} M_{\odot}}\bigg) = 
0.49^{+0.06}_{-0.05} \bigg(\frac{M_{b}}{10^{11} M_{\odot}}\bigg)^{1.17 \pm 0.08}. 
\end{eqnarray}

\cite{1996AJ....111.1889B} introduced and calculated Nuker profiles for 57 early type galaxies from HST data. This profile is described by two power laws and matches with the observational profiles very well. Instead of conventional structural parameters such as core radius and central surface brightness, new parameters like the break radius $r_{b}$, and surface brightness, $\mu_{b}$, at that radius were used. Another parameter $\alpha$ describes the sharpness of the break and they have calculated these parameters by applying $\chi^{2}$ minimization technique to the mean surface brightness profiles of the early type galaxies. \cite{1997AJ....114.1771F} have analyzed 61 elliptical galaxies and spiral bulges from HST data and derived the parameters like $r_b$, the intensity at that radius, $I_b, \sigma$ and $L$.  \cite{2004ApJ...600..149W} and \cite{2016MNRAS.455..859S} used these results in their spherical galaxy model for deriving the distribution function while we use it to derive the empirical $M-\sigma$ relation.

In this paper, we describe a theoretical model for calculating line of sight velocity dispersion for spherical systems and thereby derive the $M_{\bullet}- \sigma$ relation. In \S 2, we discuss the nexus between the $M_{\bullet} - \sigma$ relation and the power law mass density index analytically motivating the theoretical models of power law galaxies. In the \S 3, we have extended the model to the case of Nuker intensity profile, which is much more generalized than the special case of a single power law profile. Using parameters derived from the observational profiles for 12 galaxies we have determined the $M_{\bullet}- \sigma$ relation and the $M_{b}$ - $M_{\bullet}$ relation for the proportionality case from $\chi^{2}$ analysis. We discuss our results in \S 5 and present our conclusions in \S 6.

\section{Connection of $M_{\bullet} - \sigma$ relation with power law mass density of galaxies \label{s1}}
If $M_{\bullet}$ is proportional to $M_{b}$, then eq. (\ref{m_sigma_rel}) can be written as
\begin{eqnarray}
f M_{b} = k \sigma^p.\label{th_pl} 
\end{eqnarray}
The total mass scales as $\rho r^{3}$, where, $\rho$ is the mass density of the galaxy and $r$ is the distance from the center of the galaxy; similarly $\sigma$ scales as $\sqrt{\rho r^{2}}$. Therefore, from the eq. (\ref{th_pl}), it can be seen that
\begin{eqnarray}\nonumber
\rho r^{3} \propto \rho^{\frac{p}{2}} r^{p}\\
\Rightarrow \rho \propto r^{\frac{2p - 6}{2 - p}}.
\label{th_pl_f}
\end{eqnarray}
From the above relation it can be infirmed that the density follows a single power law so that
\begin{eqnarray}
\gamma = \frac{2p - 6}{2 - p};{\rm ~or ~ equivalently} ~ p = \frac{2\gamma + 6}{2 + \gamma},
\label{pl_slope}
\end{eqnarray}
where $\gamma$ is the power law index. Taking typical observational values for $p$, we find $\gamma$  = $0.75 - 1.4$ giving $p$ = 3.6 - 5.3. For a single power law profile given by 
\begin{eqnarray}
\rho (r) = \rho_{0} \bigg(\frac{r}{r_{0}}\bigg) ^{-\gamma},
\end{eqnarray}
we use Poisson's equation to calculate the stellar potential of the system
\begin{eqnarray}
\nabla^{2} \Phi = 4\pi G\rho,
\end{eqnarray}
to find a stellar potential of the form
\begin{eqnarray}
\psi_{\star} (r) = \frac{4\pi G\rho_{0} r_{0}^{\gamma} r_{h}^{2 - \gamma}}{(2 - \gamma) (3 - \gamma)} \bigg[1 - \bigg(\frac{r}{r_{h}}\bigg)^{2 - \gamma} \bigg].
\end{eqnarray}
The total mass of stars contained within $r_{h}$ is given by
\begin{eqnarray}\nonumber
M_{\star} (r < r_{h}) = \int_{0}^{r_{h}} \rho (r) 4\pi r^{2} \diff r\\
= 4\pi \rho_{0} r_{0}^{\gamma} \int_{0}^{r_{h}} r^{2 - \gamma} \diff r = 4\pi \rho_{0} r_{0}^{\gamma} \frac{r_{h}^{3 - \gamma}}{3 - \gamma} = 2M_{\bullet}, \label{def_rh}
\end{eqnarray}
where, $\rho_{0} r_{0}^{\gamma} = \frac{(3 - \gamma)}{2\pi} M_{\bullet} r_{h}^{\gamma - 3}$, so that the stellar potential takes the form  
\begin{eqnarray}
\psi_{\star} (r) = \frac{2}{2 - \gamma} \frac{G M_{\bullet}}{r_{h}}\bigg[1- \bigg(\frac{r}{r_{h}}\bigg)^{2 - \gamma}\bigg],
\end{eqnarray}
and the total potential is given by
\begin{eqnarray}
\psi (r) = \psi_{\star} (r) + \frac{GM_{\bullet}}{r} + \psi_{c},
\end{eqnarray}
where, $\psi_{c}$ is a constant which ensures that $\psi(r)$ asymptotes to zero. We normalize the total potential in units of $G M_{\bullet}/r_{h}$ so that
\begin{eqnarray}
\psi = \frac{1}{r_{\star}} + \frac{2}{2-\gamma} (1 - (r_{\star})^{2-\gamma}) + \psi_{0} = x +  \frac{2}{2-\gamma} (1 - (x)^{\gamma - 2}) + \psi_{0},
\end{eqnarray}
where, 
\begin{eqnarray}
r_{\star} = \frac{r}{r_{h}}, x = \frac{1}{r_{\star}} ,\psi_{0} = \frac{\psi_{c}}{\frac{G M_{\bullet}}{r_{h}}}.
\end{eqnarray}
Next, we calculate the distribution function from Eddington's formula as
\begin{eqnarray}
f(\epsilon) = \frac{1}{\sqrt{8}\pi^{2}m_{\star}} \frac{\diff}{\diff\epsilon} \int_{0}^{\epsilon} \frac{\diff\rho}{\diff\psi} \frac{\diff\psi}{\sqrt{\psi - \epsilon}}, \label{Edd_frmla}
\end{eqnarray}

where, $m_{\star}$ is the stellar mass which results in
\begin{eqnarray}\nonumber
f(\epsilon) = \frac{\gamma(3 - \gamma)}{4\sqrt{2} \pi^{3}} \frac{1}{m_{\star}}\frac{1}{G^{3}M_{\bullet}^{2}} g(\epsilon),
\end{eqnarray}
where, 
\begin{eqnarray}
g(\epsilon) = \frac{\diff}{\diff \epsilon} \int_{x_{1}}^{x_{2}} \frac{x^{\gamma - 1}}{\sqrt{\epsilon - x - \frac{2}{2 - \gamma} (1 - x^{\gamma - 2})}} dx ,
\end{eqnarray}
and $x_{1}$ and $x_{2}$ are the roots of the equations $\psi(x) = 0$ and $\psi(x) = \epsilon$ respectively.

\begin{figure}[H]
\begin{center}
\includegraphics[scale=0.35]{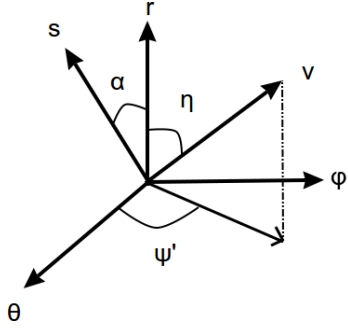}
\caption{The velocity vector {\bf v}, LOS direction $\hat{s}$ and the definition of the angles $\alpha$, $\eta$ and $\psi^{\prime}$ are sketched.}
\label{fig:axes}
\end{center}
\end{figure}

The LOS velocity dispersion is given by \citep{2008gady.book.....B} 
\begin{eqnarray}
\sigma_{||}^{2} = \frac{\int \diff x_{||} \diff^{3}{\bf v} v_{||}^{2} f(x, v)}{\int \diff x_{||} \diff^{3}{\bf v} f(x, v)}. \label{los}
\end{eqnarray}
We use $\sigma$ in place of $\sigma_{||}$ for the rest of the paper, consider the system to be spherical, and use polar coordinates in velocity space as (see Fig. \ref{fig:axes}),
\begin{eqnarray}
v_{r} = v \cos \eta, v_{\theta} = v \sin \eta \cos \psi^{\prime}, v_{\phi} = v \sin \eta \sin \psi^{\prime}.
\end{eqnarray}
We take the LOS (line of sight) direction to be an arbitrary direction, $\hat{s}$, which lies in the $r - \theta$ plane making an angle $\alpha$ with $\hat{r}$ axis so that
\begin{eqnarray}
\hat{s} = \cos\alpha \hat{r} + \sin\alpha\hat{\theta}.
\end{eqnarray}

The projected velocity in this plane of LOS is given by
\begin{eqnarray}
{\bf v}.\hat{s} = v_{||} = v\cos\eta\cos\alpha + v\sin\eta\cos\psi^{\prime}\sin\alpha.
\end{eqnarray}
The distance along the LOS is now $x_{||} = r\cos\alpha$ where the perpendicular distance is $x_{\perp} = \omega = r\sin\alpha$, and
\begin{eqnarray}
x_{||} = \sqrt{r^{2} - r^{2}\sin^{2}\alpha} = \sqrt{r^{2} - \omega^{2}},
\end{eqnarray}
where, $r^{2}$ varies from $\omega^{2}$ to $\infty$.
We find the denominator $D$ and the numerator $N$ of the LOS velocity dispersion (eq. (\ref{los})) separately as

\begin{eqnarray}\nonumber
D_{1} = \int \diff x_{||} \diff ^{3}{\bf v} f(x, v)
\end{eqnarray}
\begin{eqnarray}
= \frac{1}{2} \int_{r^{2} = \omega^{2}}^{\infty} \frac{\diff (r^{2})}{\sqrt{r^{2} - \omega^{2}}} \int_{v = 0}^{\sqrt{2\psi}} \int_{\eta = 0}^{\pi} \int_{\psi^{\prime} = 0}^{2\pi} v^{2} \diff v \sin\eta \diff \eta \diff \psi^{\prime} f(\varepsilon),
\end{eqnarray}

which after substituting $u = \omega^{2} / r^{2}$ and $v^{2} = 2(\psi - \varepsilon)$ reduces to 
\begin{eqnarray}
 D_{1} = \pi \omega J_{0} \int_{0}^{1} \frac{\diff u}{u^{2}\sqrt{\frac{1}{u} - 1}} \int_{0}^{\psi} \diff \varepsilon(2(\psi - \varepsilon))^{\frac{1}{2}} f(\varepsilon),
\end{eqnarray}
 where, $J_{0} = \int_{0}^{\pi} \sin\eta \diff \eta = 2.$ Now,

\begin{eqnarray}\nonumber
 N_{1} = \int \diff x_{||} \diff ^{3}{\bf v} v_{||}^{2} f(x, v)
\end{eqnarray}
\begin{eqnarray}
= \frac{1}{2} \int_{r^{2} = \omega^{2}}^{\infty} \frac{\diff (r^{2})}{\sqrt{r^{2} - \omega^{2}}} \int_{v = 0}^{\sqrt{2\psi}} \int_{\eta = 0}^{\pi} \int_{\psi^{\prime} = 0}^{2\pi} v^{2} \diff v \sin\eta \diff \eta \diff \psi^{\prime}
v^{2} \bigg( \cos\eta \frac{\sqrt{r^{2} - \omega^{2}}}{\sqrt{r^{2}}}  + \sin\eta \cos\psi^{\prime} \sqrt{\frac{\omega^{2}}{r^{2}}}\bigg)^{2}f(\varepsilon).
\end{eqnarray}

\begin{figure}[H]
\begin{center}
\includegraphics[scale=0.34]{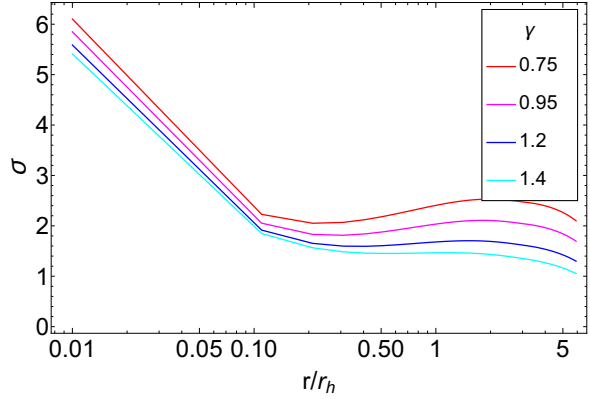}
\caption{The dimensionless $\sigma$ is plotted against projected $r / r_{h}$ for various power law indices $\gamma$. \label{fig:plsigma}}
\end{center}

\end{figure}

Similarly, with the same substitutions, $N_{1}$ reduces to

\begin{eqnarray}
 N_{1} = \omega 2^{\frac{1}{2}} \bigg[2\pi J_1 \int_{0}^{1} \frac{\diff u \sqrt{\frac{1}{u} - 1}}{u} \int_{0}^{\psi} \diff \varepsilon (\psi - \varepsilon)^{\frac{3}{2}} f(\varepsilon) + J_2 J_3 \int_{0}^{1} \frac{\diff u}{u \sqrt{\frac{1}{u} - 1}} \int_{0}^{\psi} \diff \varepsilon (\psi - \varepsilon)^{\frac{3}{2}} f(\varepsilon)\bigg],
\end{eqnarray}
where, 
\begin{eqnarray}\nonumber
J_1 \int_{0}^{\pi} \sin\eta \cos^{2}\eta \diff \eta = \frac{2}{3}, J_2 = \int_{0}^{\pi} \sin^{3}\eta \diff \eta = \frac{4}{3}, J_3 \int_{0}^{2\pi} \cos^{2}\psi^{\prime} \diff \psi^{\prime} = \pi. 
\end{eqnarray}
The dimensionless LOS velocity dispersion given by $\sigma = \sqrt{\frac{N_{1}}{D_{1}}}$ for power law galaxies is shown in Fig. \ref{fig:plsigma}, where we can see that the velocity dispersion is flattening out as we move outwards from the center of the galaxy. Near the center of the galaxy where the SMBH potential dominates $\sigma \propto 1 / \sqrt{r}$. Later it flattens out because of the dominance of the stellar potential. By finding $\sigma$ at any radius one can verify the $M_{\bullet} - \sigma$ relation if $M_{\bullet}$ is known.

From the definition, eq. (\ref{def_rh}), $r_{h}$ for a single power law galaxy can be written as
\begin{eqnarray}
r_{h} = \bigg(\rho_{0} r_{0}^{\gamma} \frac{2\pi}{3 - \gamma} \frac{1}{M_{\bullet}}\bigg)^{\frac{1}{\gamma - 3}}.
\end{eqnarray}
 
The total mass out to the bulge can be calculated to be
\begin{eqnarray}
\int_{0}^{r_{s}}\rho_{0}\bigg(\frac{r_{0}}{r}\bigg)^{\gamma} 4\pi r^{2} \diff r = M_{s},
\end{eqnarray},
where, $r_{s}$ is the radius of the central bulge.
Therefore, $\rho_{0} r_{0}^{\gamma}$ can be written as, 
\begin{eqnarray}
\rho_{0} r_{0}^{\gamma} = \frac{M_{s}}{\int_{0}^{r_{s}}4\pi r^{2-\gamma} \diff r} = \frac{M_{s}(3 - \gamma)}{4\pi r_{s}^{3 - \gamma}}.
\end{eqnarray}

\begin{figure}
\begin{center}
\subfigure[\label{fig:plslint_a}]{\includegraphics[scale=0.48]{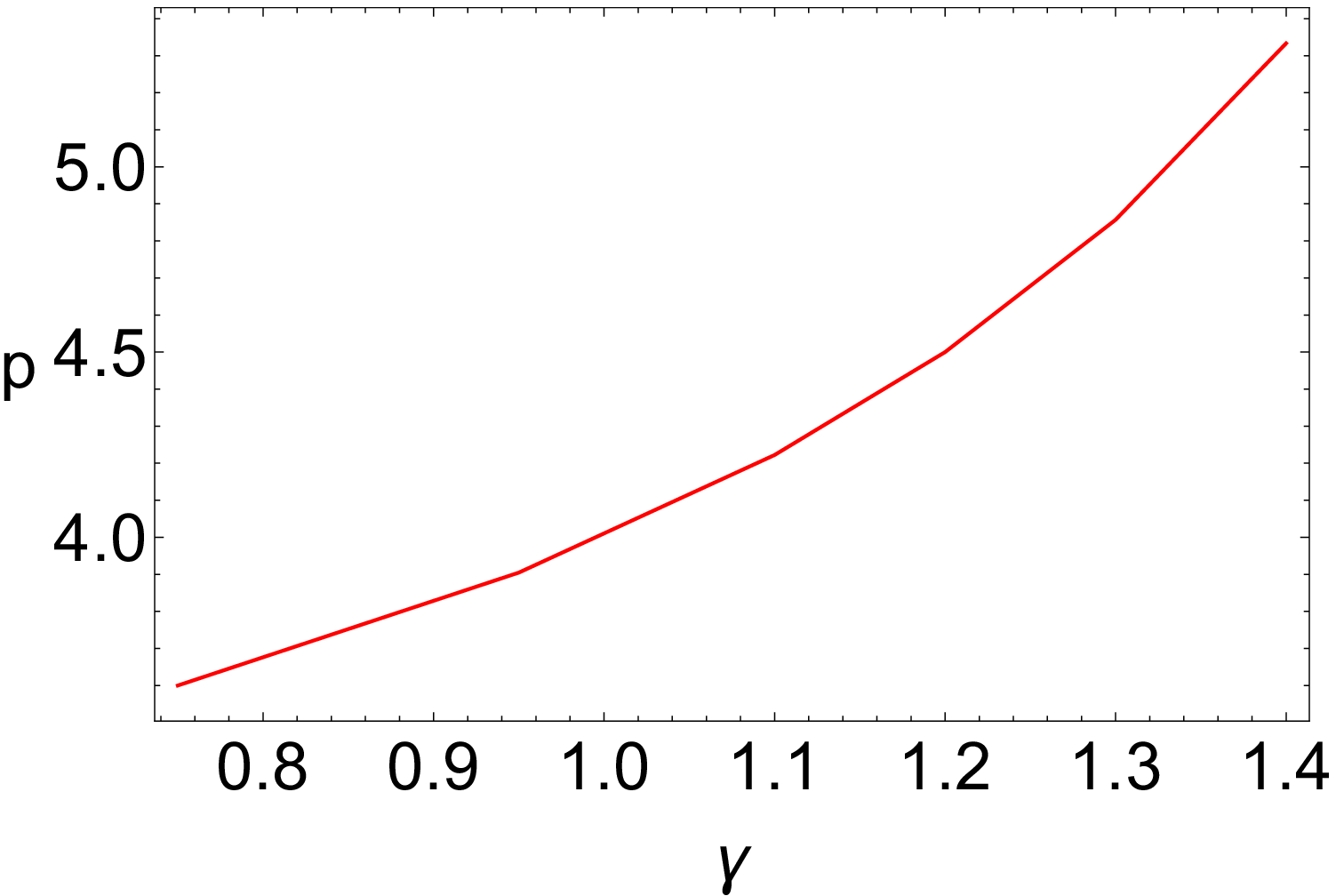}}
\subfigure[\label{fig:plslint_b}]{\includegraphics[scale=0.25]{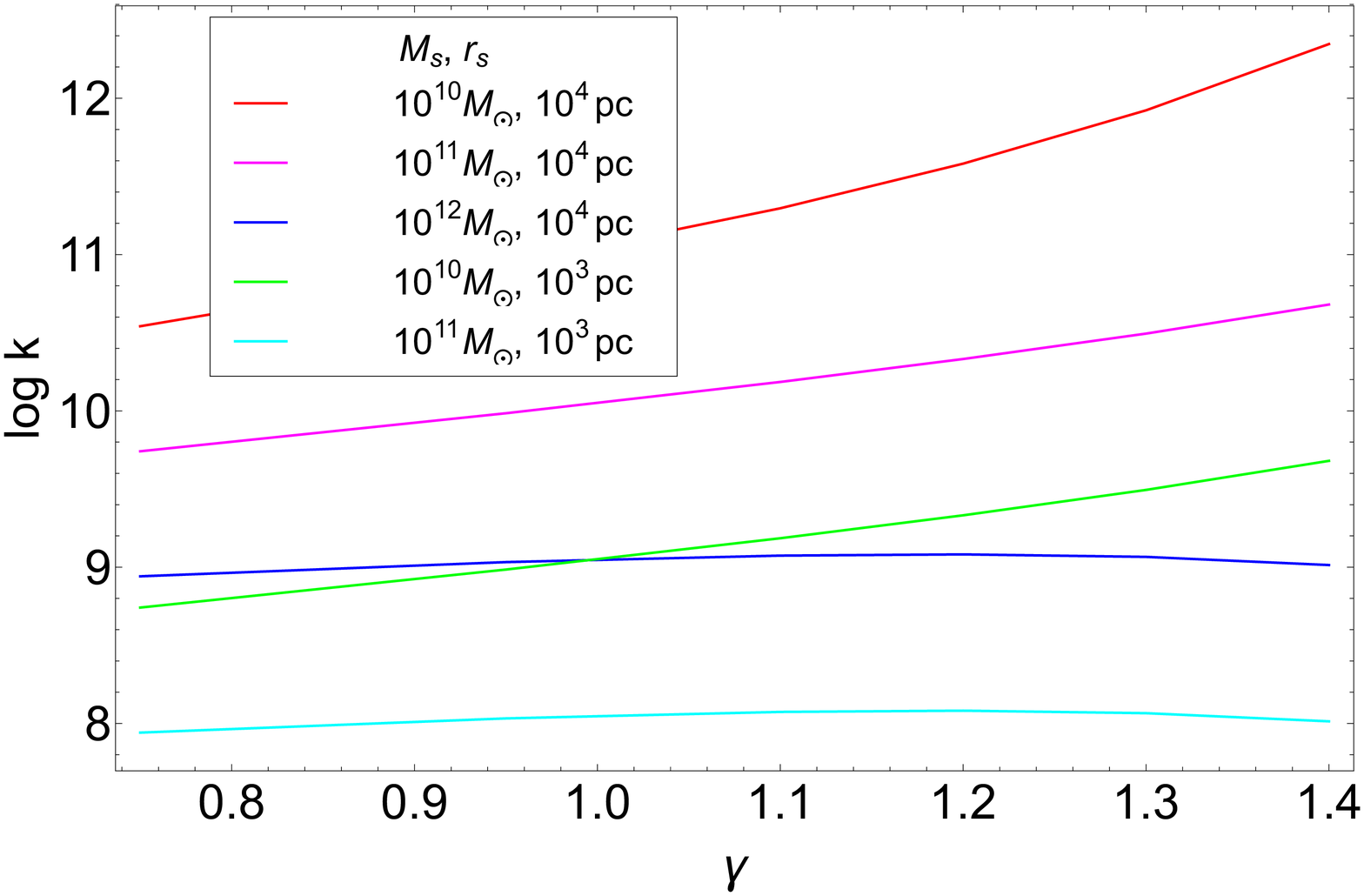}}
\subfigure[\label{pl_cont2_a}]{\includegraphics[scale=0.35]{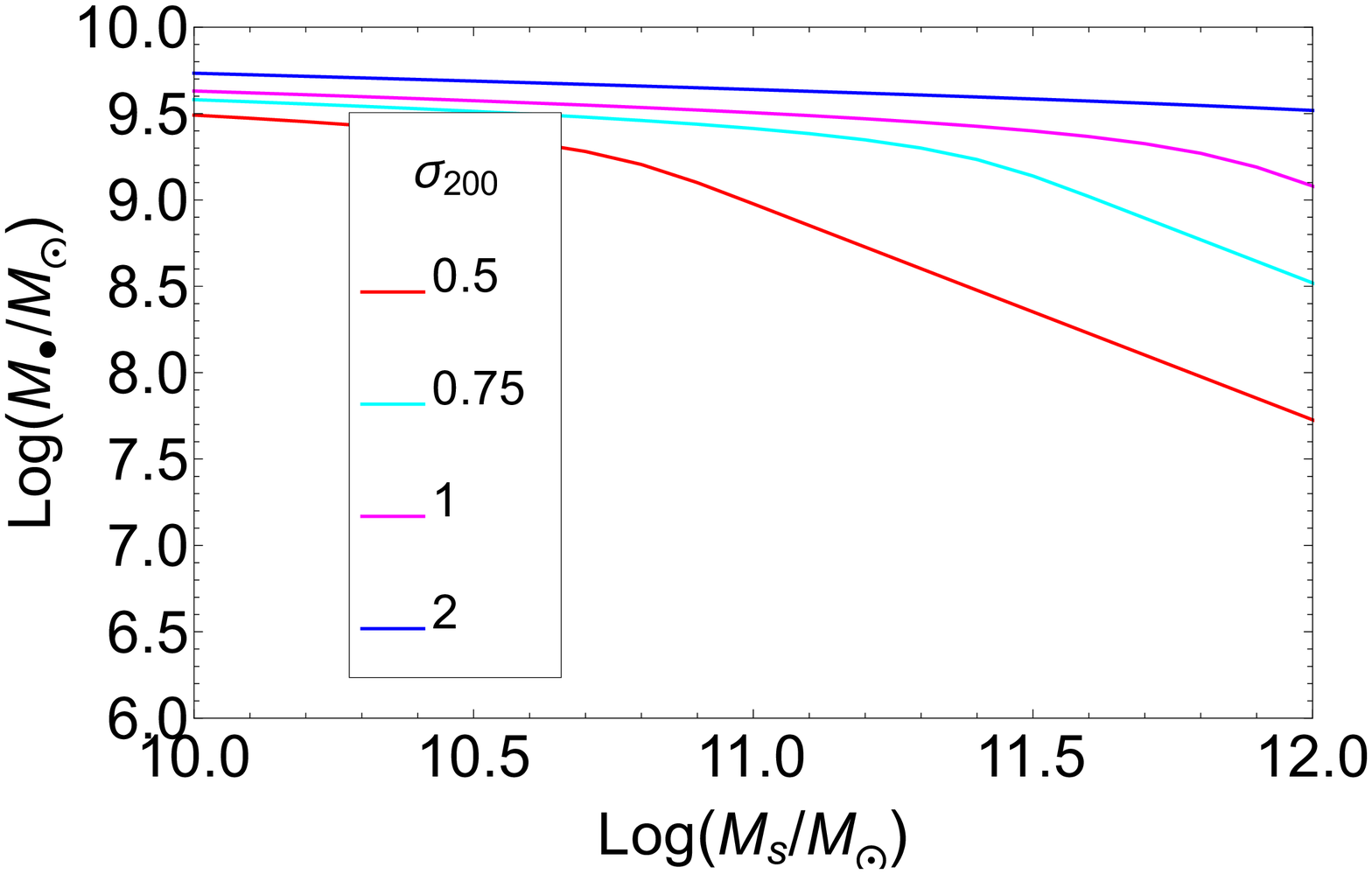}}
\subfigure[\label{pl_cont2_b}]{\includegraphics[scale=0.35]{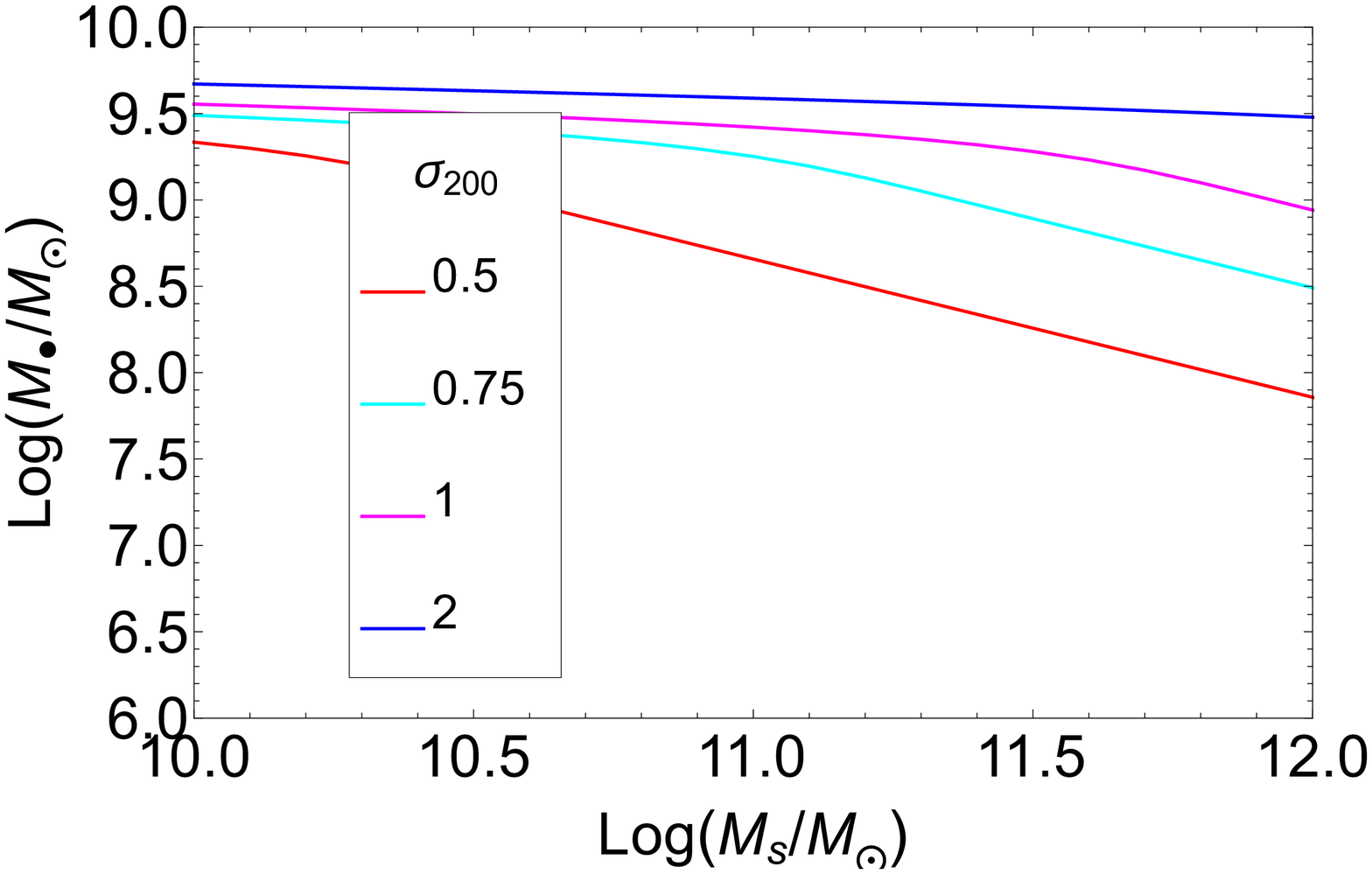}}
\end{center}
\caption{Plot of $p(\gamma)$ (left) and plot of $\log k (M_{s})$ (right) for different values of $\gamma$ for different $M_{s} -r_{s}$ combinations (up). Contour plot of $\sigma_{200}$ at 3$r_{h}$ for different power laws ($\gamma = 1.2$ (left) and $\gamma = 0.75$ (right)) by varying $M_{s}$ for different values of  $M_{\bullet}$ for a fixed $r_{s}$ = $10^{4}$ pc (down).}
\label{pl_cont2}
\end{figure}

For a range of black hole mass ($M_{\bullet}$ = $10^{6}$ to $10^{9}$ $M_{\odot}$) our calculated $\sigma(M_{s}, \gamma)$, $\log k(M_{s}, \gamma)$ and $p(\gamma)$ (which is observationally within 4 - 5) are shown in Fig. \ref{pl_cont2}, where, $M_{s}$ varies from $10^{10}$ to $10^{12}$ $M_{\odot}$, $r_{s}$ varies from 1 - 10 kpc and $\gamma$ varies from 0.75 to 1.5. We can see that for a fixed value of $\gamma$, $p$ is independent of the value of $M_{s}$. The range of $p$ we find is 3.6 - 5.3, which agrees well with the observations. Fig. \ref{fig:plslint_b} shows a plot of $\log k (M_{s}, \gamma)$; a change in $M_{s}$ for a fixed value of $\gamma$ affects the intercept though the slope is unchanged. To explain the nature of these plots we write eq. (\ref{m_sigma_rel}) as
\begin{eqnarray}
M_{\bullet} = \bigg(\frac{M_{s}}{2}\bigg)^{\frac{1}{\gamma - 2}} r_{s}^{\frac{\gamma - 3}{\gamma - 2}} (\sigma_{h}^{2} G)^{-\frac{\gamma - 3}{\gamma - 2}} \sigma^{\frac{2(\gamma - 3)}{\gamma - 2}}, \label{th_int}
\end{eqnarray}
where, $\sigma_{h}$ is the value of dimensionless $\sigma$ at 3$r_{h}$.
From the eq. (\ref{th_int}) we see that the constant $k$ depends on $\gamma$, $r_{s}$ and $M_{s}$, but the index, $p$ of the $M_{\bullet} - \sigma$ relation depends only on $\gamma$ which clearly explains the nature of the plots in Fig \ref{fig:plslint_a} and \ref{fig:plslint_b}.



The contour plot of $\sigma_{200}$ (see Fig. \ref{pl_cont2}) at 3$r_{h}$ for different power laws by varying $M_{s}$ and $M_{\bullet}$ for a fixed $r_{s}$ = $10^{4}$ pc is shown in Fig. \ref{pl_cont2}. By selecting a physical and observed range for $\sigma$, one can obtain the allowed $M_{\bullet} - M_{s}$ combinations for those systems. 

\section{Spherical galaxies following Nuker profile of intensity \label{s2}}
\begin{figure}[H]
\begin{center}
\subfigure[\label{fig:ngc_3115_1_a}]{\includegraphics[scale=0.25]{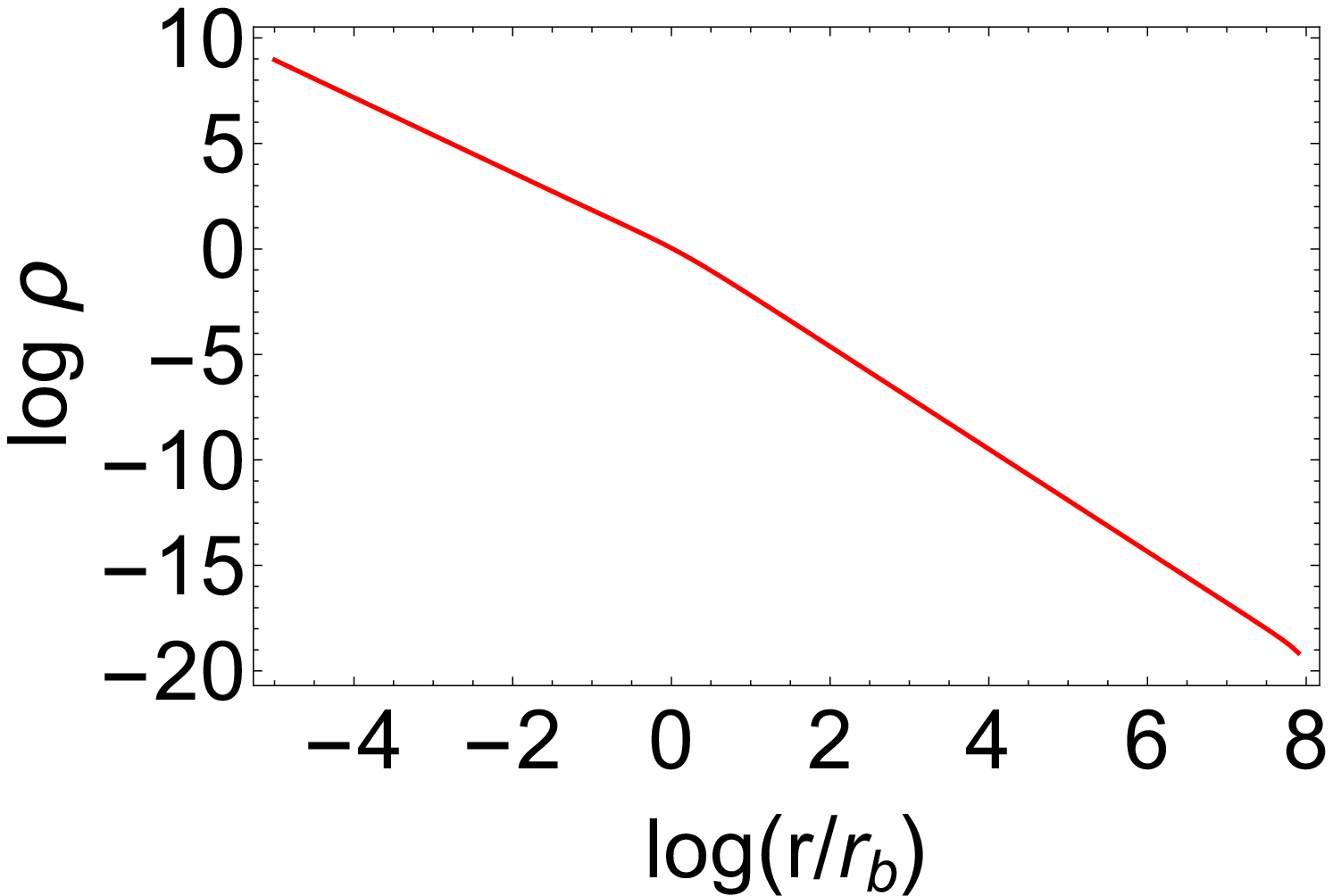}}
\subfigure[\label{fig:ngc_3115_1_b}]{\includegraphics[scale=0.25]{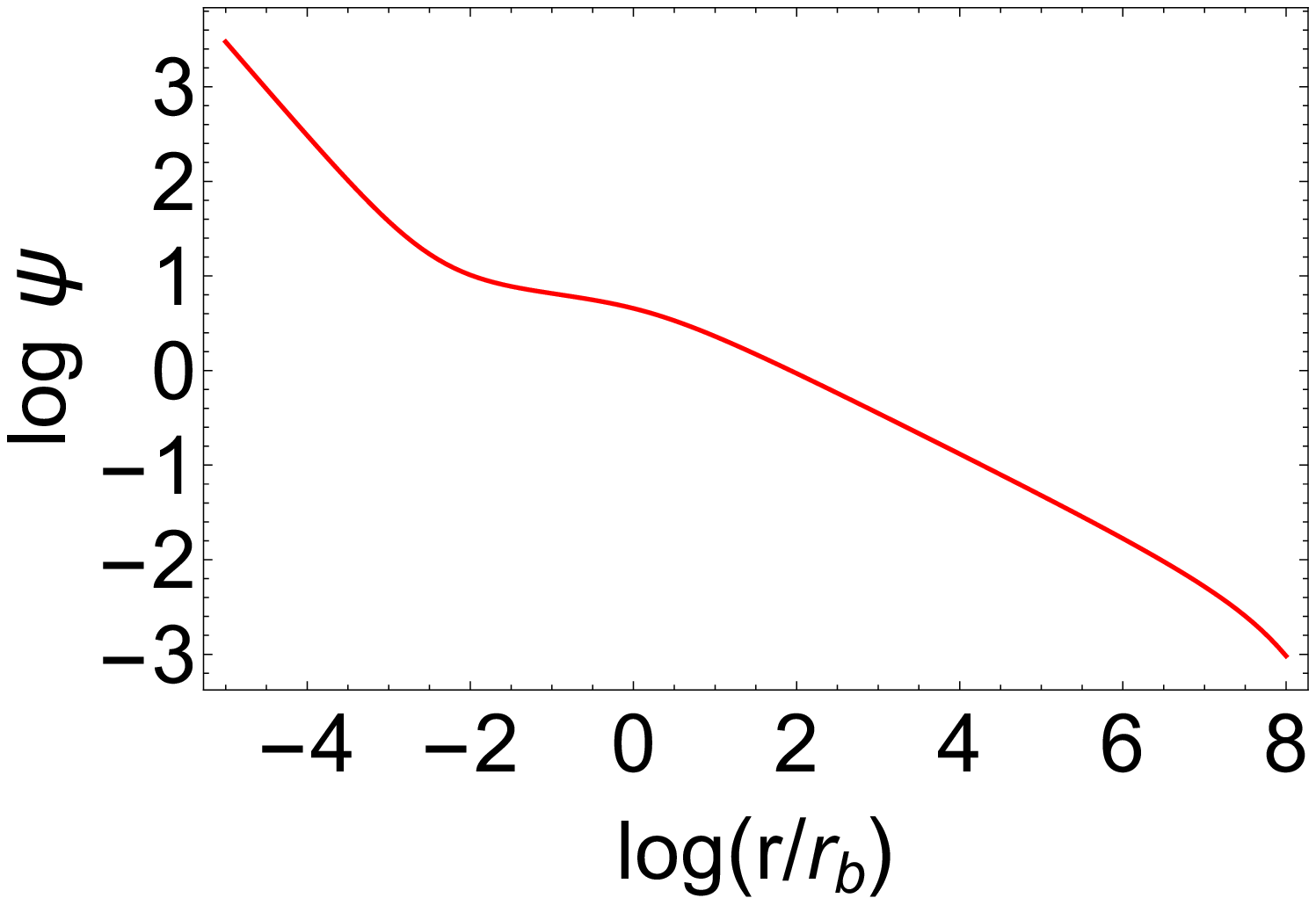}}
\subfigure[\label{fig:ngc_3115_a}]{\includegraphics[scale=0.25]{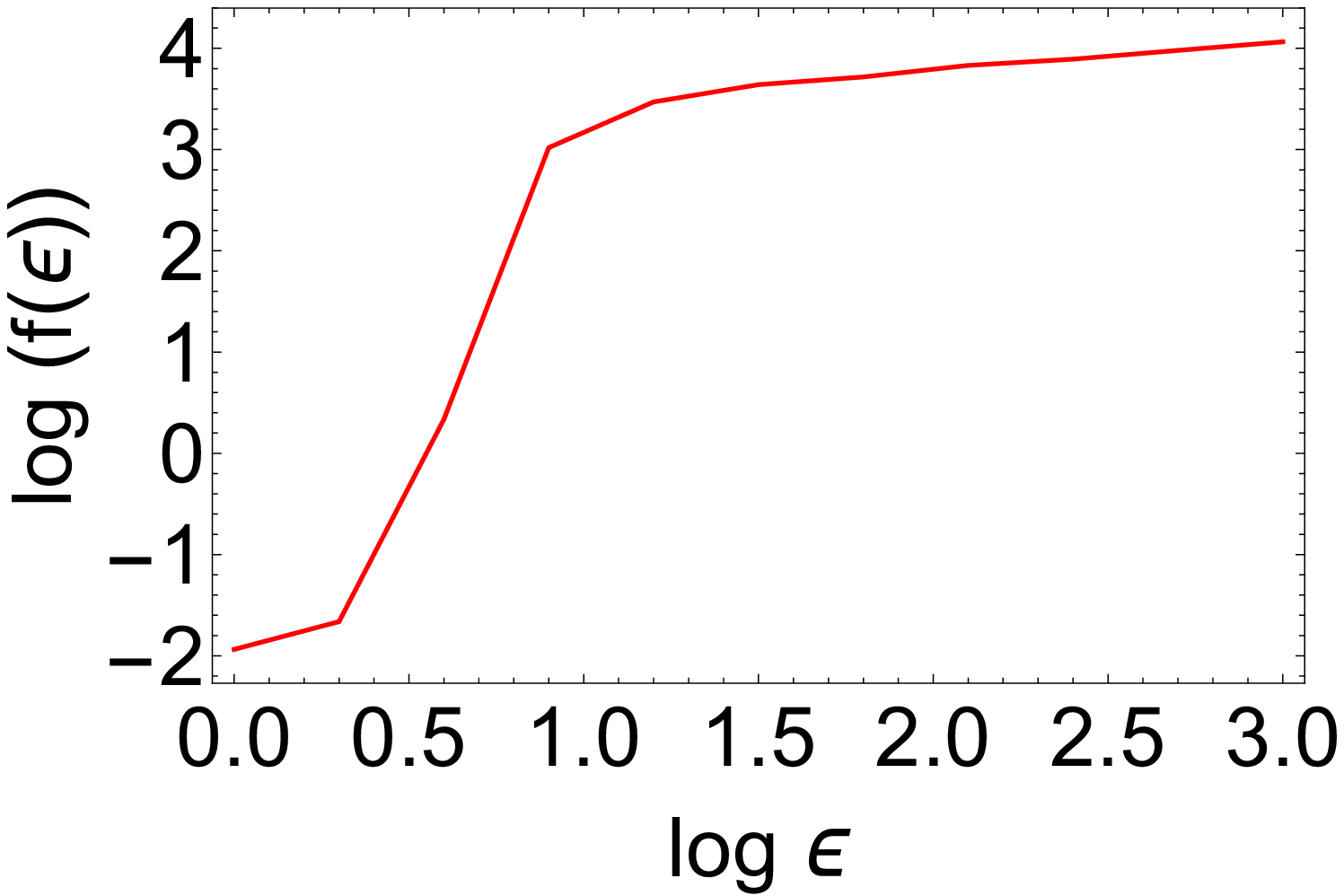}}
\subfigure[\label{fig:ngc_3115_b}]{\includegraphics[scale=0.2]{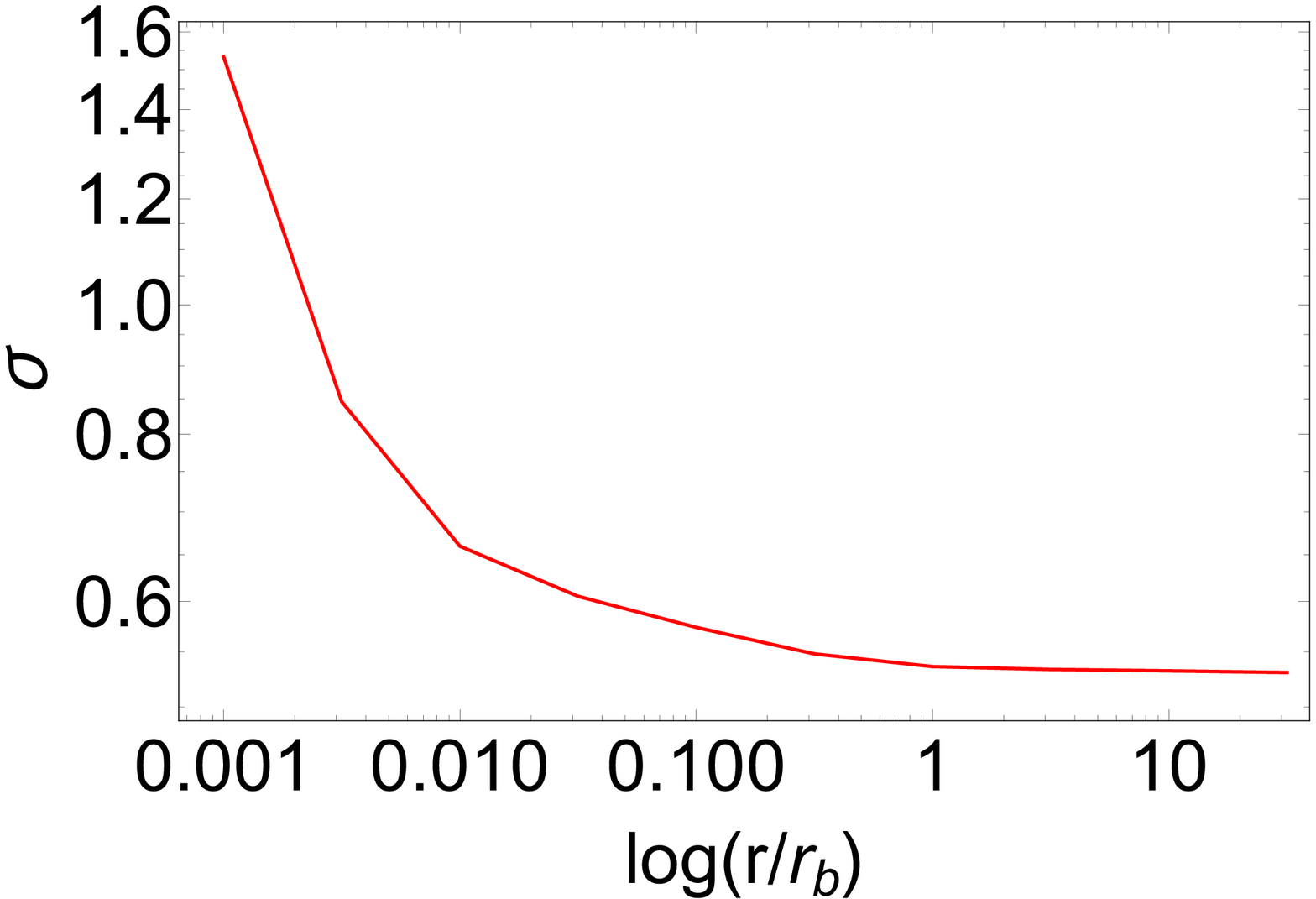}}
\end{center}
\caption{The density, total potential plots, distribution function and velocity dispersion plots from left to right for $f$ = 0.0012 for NGC 3115.}
\label{fig:ngc_3115_1}
\end{figure}

The Nuker profile used to fit the observational luminosity data is given by,
\begin{eqnarray}
I(\xi) = I_{b}2^{\frac{(\beta - \Gamma)}{\alpha}}\xi^{-\Gamma}(1 + \xi^{\alpha})^{-\frac{(\beta - \Gamma)}{\alpha}},
\end{eqnarray}
where, $\xi = \frac{R}{r_{b}}$, $r_{b}$ is the break radius and $I_{b}$ is the intensity at the break radius, $\Gamma$ is the inner slope and the outer slope is $\beta$. The visual mass - to - light ratio is denoted by $\Upsilon_{v}$ (assuming $H_{0}$ = 80 km s$^{-1}$ Mpc$^{-1}$) and $\mu_{b}$ is the surface brightness in visual magnitudes arcsec$^{-2}$ at $r_{b}$. The quantity $\mu$ represents the apparent magnitude of the equivalent total light observed in a square arcsec at different points in the distribution and it can be related to the
physical surface brightness profile through \citep{1998gaas.book.....B} :
\begin{eqnarray}
\mu = -2.5 \log I + C,
\end{eqnarray}
where $C$ is a constant. If the intensity is measured in units of $L_{\odot} {\rm pc^{-2}}$, then the constant can be calculated from the distance modulus formula and it is given as
\begin{eqnarray}
C = - 5 \log_{10}(\delta\theta) + M_{\odot}^{abs} -5,
\end{eqnarray}
where $\delta\theta$ is 1" = $\frac{1}{206265}$ radians and solar absolute magnitude, $M_{\odot}^{abs}$ is 4.83 so that $I_{b}(\mu_{b})$ can be calculated. The stellar mass density profile was computed via Abel's inversion equation as\\
\begin{eqnarray}
\rho(r) = \Upsilon_{v}j(r) = -\frac{\Upsilon_{v}}{\pi}\int_{r}^{\infty}\frac{\diff I}{\diff R}\frac{\diff R}{\sqrt{R^2 - r^2}},
\end{eqnarray}
where, $j(r)$ is the luminosity density. The stellar potential $\psi_{*}$ is calculated from the stellar mass density calculated above as shown in Fig. \ref{fig:ngc_3115_1_a}
\begin{eqnarray}
\psi_{*}(r) = \frac{4\pi G}{r}\int_{0}^{r}\rho(r^{\prime}) r^{\prime 2} \diff r^{\prime} + 4\pi G\int_{r}^{\infty}\rho(r^{\prime})r^{\prime} \diff r^{\prime}.
\end{eqnarray}

As before the gravitational potential $\psi(r) = -\Phi(r)$ is the total potential given by (see Fig. \ref{fig:ngc_3115_1_b})
\begin{eqnarray}
\psi(r) = \psi_{*}(r) + \frac{GM_{BH}}{r}.
\end{eqnarray}

The density follows the same profile (double power law) as intensity as shown in Fig. \ref{fig:ngc_3115_1_a}, where we see the total potential (see Fig. \ref{fig:ngc_3115_1_b}) is dominated by SMBH potential at the inner radii and is dominated by the stellar potential as we move outwards from the center. Again we use the Eddington's formula, eq.(\ref{Edd_frmla}) to calculate $f(\epsilon)$ shown in Fig. \ref{fig:ngc_3115_a}. The denominator of the LOS velocity dispersion can be written as
\begin{eqnarray}
D_{2}(\alpha, \beta, \Gamma, \Upsilon, r_{b}, \mu_{b}, L, f) = \int \diff x_{||} \diff^{3} {\bf v} f(x, v) \nonumber
\end{eqnarray}
\begin{eqnarray}
= \int_{r = \omega}^{\infty} \frac{r \diff r}{\sqrt{r^{2} - \omega^{2}}} \int_{v = 0}^{\sqrt{2\psi}} \int_{\eta = 0}^{\pi} \int_{\psi^{\prime} = 0}^{2\pi} v^{2} \diff v \sin\eta \diff \eta \diff \psi^{\prime} f(\varepsilon).
\end{eqnarray}
By replacing the variable $r$ by $1/u$ the denominator can finally be written as
\begin{eqnarray}
 D_{2}(\alpha, \beta, \Gamma, \Upsilon, r_{b}, \mu_{b}, L, f) = 2^{\frac{3}{2}}2\pi J_{0} \int_{u = 0}^{1 / \omega} \frac{\diff u}{u^{2}\sqrt{1-\omega^{2}u^{2}}}
  \int_{\varepsilon = 0}^{\psi} (\psi(u) - \varepsilon)^{\frac{1}{2}} f(\varepsilon) \diff \varepsilon. 
 \end{eqnarray}

The numerator $N_{2}$ of the LOS velocity dispersion is
\begin{eqnarray}\nonumber
N_{2}(\alpha, \beta, \Gamma, \Upsilon, r_{b}, \mu_{b}, L, f) = \int \diff x_{||} \diff ^{3}{\bf v} v_{||}^{2} f(x, v)
\end{eqnarray}

\begin{eqnarray}
= \frac{1}{2} \int_{r^{2} = \omega^{2}}^{\infty} \frac{\diff (r^{2})}{\sqrt{r^{2} - \omega^{2}}} \int_{v = 0}^{\sqrt{2\psi}} \int_{\eta = 0}^{\pi} \int_{\psi^{\prime} = 0}^{2\pi} v^{2} \diff v \sin\eta \diff \eta \diff \psi^{\prime}v^{2} 
\bigg( \cos\eta \frac{\sqrt{r^{2} - \omega^{2}}}{\sqrt{r^{2}}}  + \sin\eta \cos\psi^{\prime} \sqrt{\frac{\omega^{2}}{r^{2}}}\bigg)^{2}f(\varepsilon).
\end{eqnarray}
This finally takes the form
\begin{eqnarray}
N_{2}(\alpha, \beta, \Gamma, \Upsilon, r_{b}, \mu_{b}, L, f) =2^{\frac{5}{2}} \bigg[2\pi J_1 \int_{0}^{1 / \omega} \frac{\sqrt{1 - \omega^{2}u^{2}}}{u^{2}} \int_{0}^{\psi} \diff \varepsilon  (\psi - \varepsilon)^{\frac{3}{2}} f(\varepsilon)\nonumber \\ 
+ J_2 J_3 \int_{0}^{1 / \omega} \frac{\omega^{2}}{\sqrt{1 - \omega^{2}u^{2}}} \int_{0}^{\psi} \diff \varepsilon (\psi - \varepsilon)^{\frac{3}{2}} f(\varepsilon)\bigg].
\end{eqnarray}

Using the same procedure as was done in the case of power law galaxies, we compute the LOS velocity dispersion for these galaxies shown in Fig. \ref{fig:ngc_3115_b}. The distribution function (see Fig. \ref{fig:ngc_3115_a}) increases towards the higher side of the energy value implying presence of more high energy stars. Here also the velocity dispersion plot flattens out as we move outwards from the center of the galaxy where the motion of the stars are dominated by the stellar potential.\\


To simplify our calculation we use the following scales:
\begin{eqnarray}
\rho_{s} = -\frac{\Upsilon_{v}}{\pi}\frac{I_{b}}{r_{b}},~~
\psi_{s} = 4\pi G r_{b}^{2} \rho_{s},~~
f_{s} = \frac{\rho_{s}}{\sqrt{8}\pi^{2}m_{*}\psi_{s}^{\frac{3}{2}}},
\end{eqnarray}

so that $\sigma$ is in units of $\sqrt{\psi_{s}}.$ 
In Table \ref{data_nuker} we tabulate the values of $\sigma (\sqrt{Q})$ at radius $r_{e}/8$, where, $r_{e}$ is the effective radius \citep{2000ApJ...539L...9F}, where, $Q(\alpha, \beta, \Gamma, r_{b}, \mu_{b}) = N_{2} / D_{2}$. The value of $r_{e}$ is obtained from
\begin{eqnarray}
\int_{0}^{r_{e}} I(R) 2\pi R dR = \frac{1}{2} L_{T}.
\end{eqnarray}

\begin{table}[H]
\begin{center}
\scalebox{0.75}{
\begin{tabular}{|c |c |c |c |c |c |c |c |c |c |c |c |}\hline
\# & Galaxy & $\log(\frac{r_{b}}{\rm pc})$  & $\mu_{b}$ & $\alpha$ & $\beta$ & $\Gamma$ & $\Upsilon_{v}$ ($\frac{M_{\odot}}{L_{\odot}}$) & $\log(\frac{L_{V}}{L_{\odot}})$ & $\frac{M_{b}}  {10^{10} M_{\odot}}$ & $\log(\frac{r_{e}}{\rm pc})$ & $\sigma$ (${\rm km/sec}$)\\ \hline\hline
1 & NGC 3379 & 1.92 & 16.10 & 1.59 & 1.43 & 0.18 & 6.87 & 10.15 & 0.90 & 3.17 & 230\\ \hline
2 & NGC 3377 & 0.64 & 12.85 & 1.92 & 1.33 & 0.29 & 2.88 & 9.81 & 1.86 & 3.15 & 217\\ \hline
3 & NGC 4486 & 2.75 & 17.86 & 2.82 & 1.39 & 0.25 & 17.70 & 10.88 & 134.30 & 3.76 & 433\\ \hline
4 & NGC 4551 & 2.46 & 18.83 & 2.94 & 1.23 & 0.80 & 7.25 & 9.57 & 2.69 & 3.03 & 218\\ \hline
5 & NGC 4472 & 2.25 & 16.66 & 2.08 & 1.17 & 0.04 & 9.20 & 10.96 & 83.90 & 3.75 & 542\\ \hline
6 & NGC 3115 & 2.07 & 16.17 & 1.47 & 1.43 & 0.78 & 7.14 & 10.23 & 12.12 & 3.15 & 230\\ \hline
7 & NGC 4467 & 2.38 & 19.98 & 7.52 & 2.13 & 0.98 & 6.27 & 8.75 & 0.35 & 2.81 & 108\\ \hline
8 & NGC 4365 & 2.25 & 16.77 & 2.06 & 1.27 & 0.15 & 8.40 & 10.76 & 48.34 & 3.68 & 524\\ \hline
9 & NGC 4636 & 2.38 & 17.72 & 1.64 & 1.33 & 0.13 & 10.40 & 10.60 & 41.40 & 3.77 & 354\\ \hline
10 & NGC 4889 & 2.88 & 18.01 & 2.61 & 1.35 & 0.05 & 11.20 & 11.28 & 213.4 & 4.10 & 469\\ \hline
11 & NGC 4464 & 1.95 & 17.35 & 1.64 & 1.68 & 0.88 & 4.82 & 9.22 & 0.80 & 2.70 & 157\\ \hline
12 & NGC 4697 & 2.12 & 16.93 & 24.9 & 1.04 & 0.74 & 6.78 & 10.34 & 14.83 & 3.36 & 215\\ \hline \hline
\end{tabular}
}
\caption{The first 10 columns the data used for our calculation are shown ($r_{b}$ is the break radius, $\mu_{b}$ is the surface brightness, inner slope is $\Gamma$ and the outer slope is $\beta$, sharpness of break is given by $\alpha$, $\Upsilon$ is the mass - to -light ratio, $L$ is the total luminosity, the bulge mass $M_{b}$ = $\Upsilon_{v} L$) and in the last two columns, the output values of $r_{e}$ and the LOS velocity dispersion are shown \citep{2004ApJ...600..149W}.}
\label{data_nuker}
\end{center}
\end{table}

\section{$M_{\bullet} - \sigma$ relation}

\begin{figure}[H]
\begin{center}
\subfigure[\label{fig:m_sigma_fit_a}]{\includegraphics[scale=0.3]{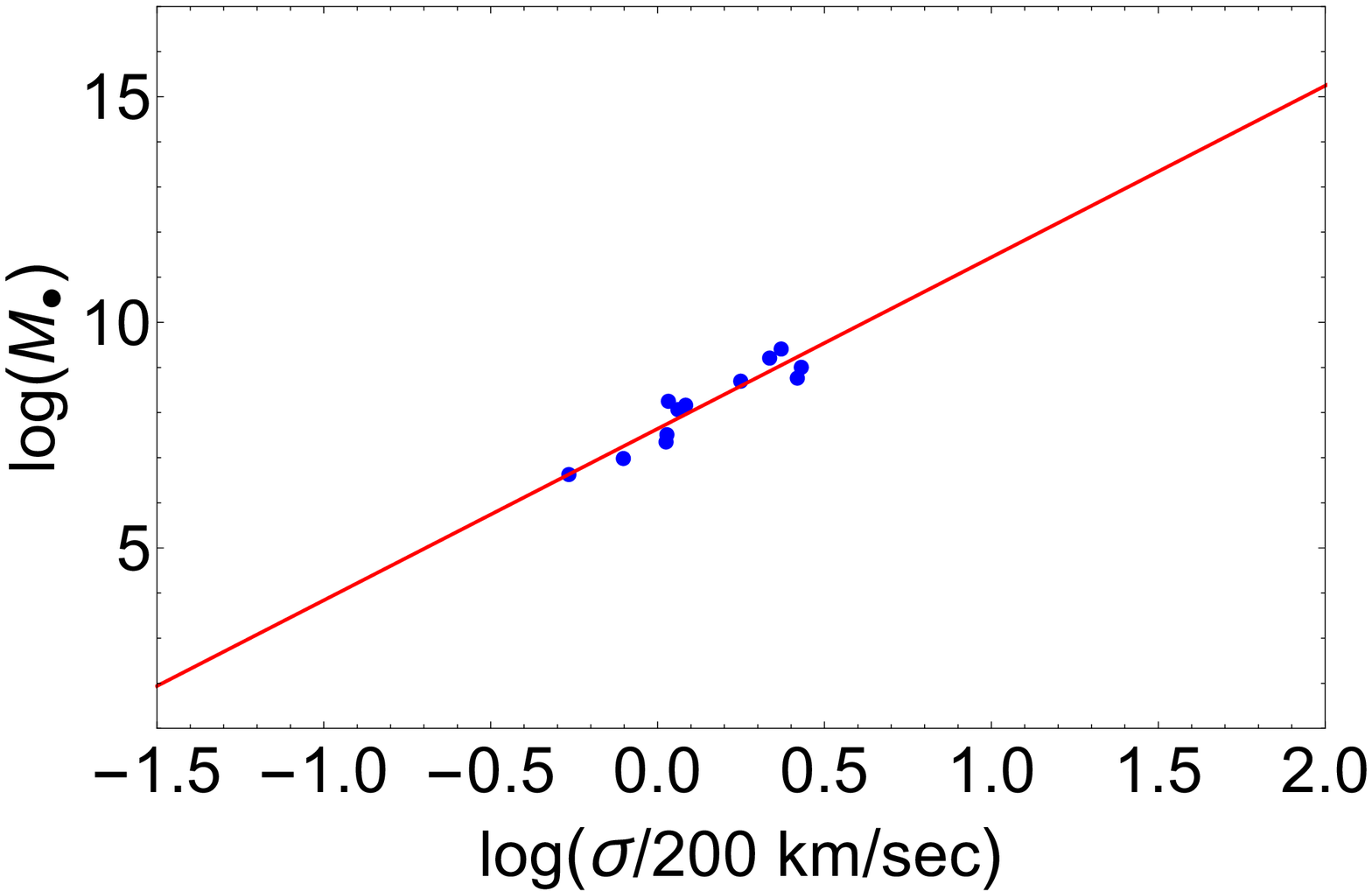}}
\subfigure[\label{fig:m_sigma_fit_b}]{\includegraphics[scale=0.3]{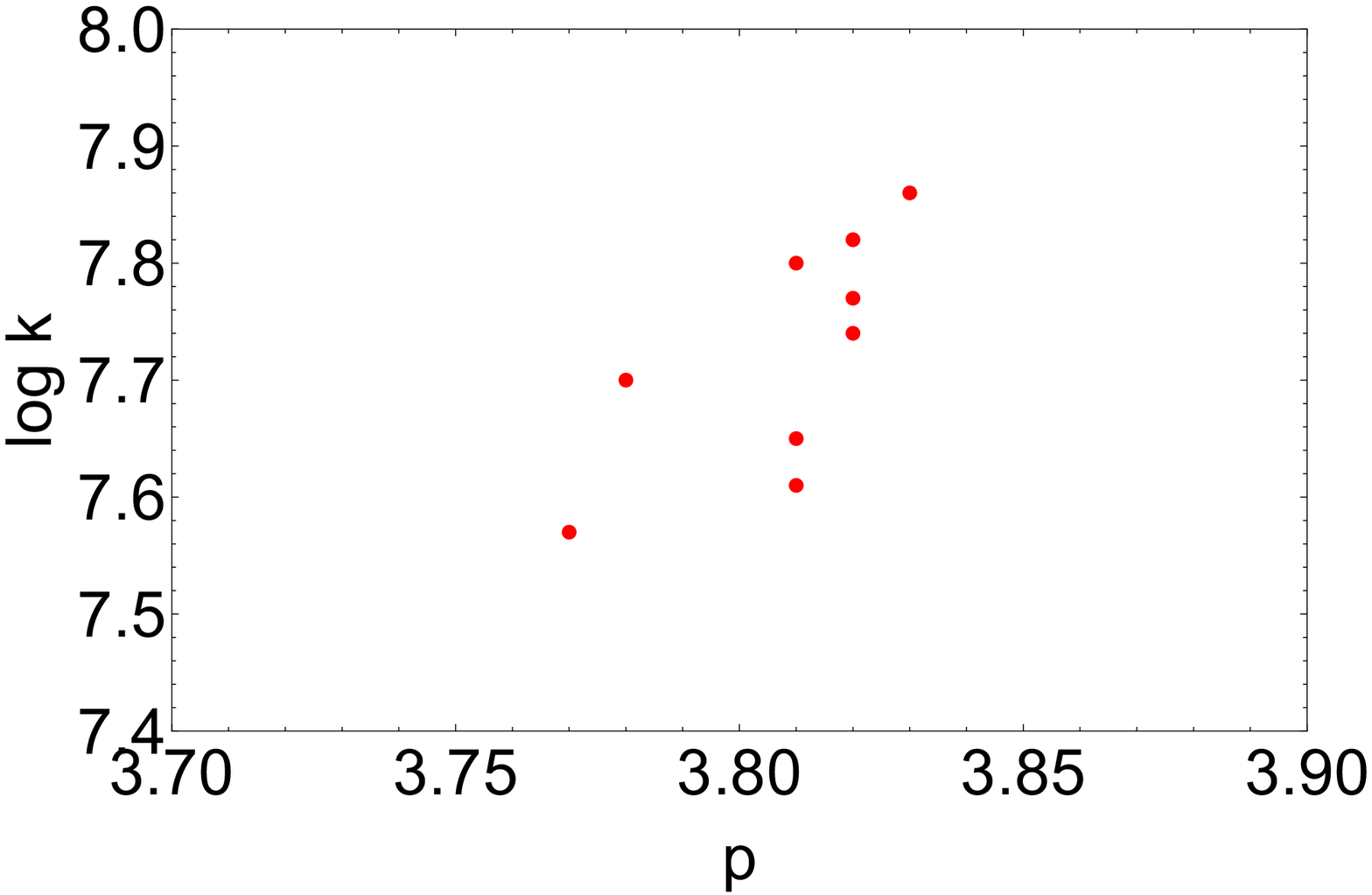}}
\end{center}
\caption{The plot of $\log \sigma$ vs $\log M_{\bullet}$ for 12 galaxies for $f = 0.0012$, $\sigma$ is in units of 200 km/sec (left) and (right) the scatter plot of $p$ and $\log k$, for different values of $f$, showing a tight range of $k$ and $p$.}
\label{fig:m_sigma_fit}
\end{figure}

\begin{figure}[H]
\begin{center}

\scalebox{0.55}{
\begin{tikzpicture}[node distance = 3cm, auto]
\tikzstyle{decision} = [diamond, draw,  
    text width=5em, text badly centered, node distance=3cm, inner sep=0pt]
\tikzstyle{block} = [rectangle, draw, 
    text width=8em, text centered, rounded corners, minimum height=1em]
\tikzstyle{line} = [draw, thick,-latex']
\tikzstyle{cloud} = [draw, ellipse,node distance=3cm,
    minimum height=4em]
\node [block,fill=white] (para) {{\bf \color{black}Data of Nuker intensity profile and data of bulge mass ($M_{b}$)}};
\node [block,fill=white, right of= para, node distance=6cm] (para1) {{\bf Stellar mass density ($\rho(r)$)}};
\node [block,fill=white, right of= para1,node distance=6cm] (para2) {{\bf Stellar potential ($(\psi_{*}$)}};

\node [block,fill=white, right of= para2,node distance=6cm] (para4) {{\bf Total potential ($\psi = \psi_{*} + \psi_{BH}$)}};
\node [block,fill=white, right of= para4,node distance=6cm] (para5) {{\bf Distribution function of stars ($f(\epsilon)$)}};
\node [block, fill=white, below of=para4, node distance=4cm] (para3) {{\bf Line of sight velocity dispersion of stars ($\sigma$)}};

\node [cloud,fill=white, below of= para2, node distance=4cm] (para6) {{\bf \color{black}$M_{\bullet} - \sigma$ relation}};

\node [block,fill=white, below of= para1, node distance=4cm] (para8) {{\bf Mass of the SMBH ($M_{\bullet}$)}};

\draw[->,thick,line width=0.5mm,draw=blue!80] (para.south) to [out=270,in=90] (para8.north);
    \draw[->,thick,line width=0.5mm,draw=blue!80] (para5.south) to [out=270,in=90] (para3.north);

\path [line] (para)--node{ }(para1);
\path [line,blue] (para1)--node{ }(para2);
\path [line,blue] (para2)--node{ }(para4);
\path [line,blue] (para4)--node{ }(para5);
\path [line,blue] (para3)--node{ }(para6);
\path [line,blue] (para8)--node{ }(para6);

 \end{tikzpicture}

}
\caption{The flowchart shows the procedure for calculating the $M_{\bullet} - \sigma$ relation from observational data. For the Nuker profile the stellar mass density is found using Abel inversion and from spherical shell structure the stellar potential is calculated. The SMBH potential is added to get the total potential, Eddington's formula is used to derive distribution function $f(\epsilon)$. The SMBH mass is calculated from the proportionality relation of $M_{b}$ and $M_{\bullet}$. The path marked by the blue lines only is followed for deriving the $M_{\bullet} - \sigma$ relation in case of single power law galaxies.}
\label{flowchart}
\end{center}
\end{figure}
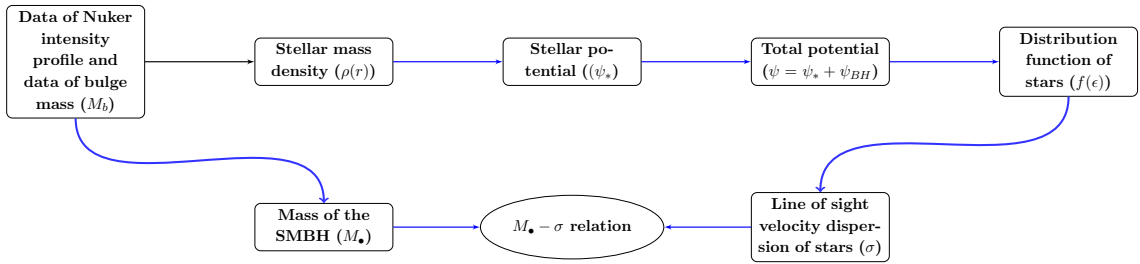

By using the data given in \cite{2004ApJ...600..149W} for elliptical galaxies as shown in Table \ref{data_nuker}, we calculate the bulge mass of those galaxies by multiplying total luminosity by the mass to light ratio as prescribed in \cite{1998AJ....115.2285M}. Using eq. (\ref{los}) we calculate the LOS velocity dispersion to get the $M_{\bullet}$ - $\sigma$ relation by fitting a straight line for 12 galaxies as shown in Fig. \ref{fig:m_sigma_fit_a} for different $f$ values. The $p$ and $\log k$ values for different values of $f$ are shown in the scatter plot (see Fig. \ref{fig:m_sigma_fit_b}). From $\chi^{2}$ minimization we have determined $p$ and $f$; the procedure used is shown in a flowchart given in Fig. \ref{flowchart}. 

The formula we used for determining $\chi^{2}$ is \citep{2006data.book.....S}
\begin{eqnarray}
\chi^{2} = \sum_{k}\bigg[\frac{(D_{k} - F_{k})^{2}}{F_{k}}\bigg], \label{chisq}
\end{eqnarray}
where a uniform prior is used. The observed values, $D_{k}$ are obtained for $M_{\bullet}$ and $\sigma$ from our calculation using the observational data and the expected value, $F_{k}$ is obtained by the best fit straight line to these points as shown in Fig. \ref{fig:m_sigma_fit_a}. The range of $f$ and $p$ has been taken from previous determinations as well as observational values ($f$ = 0.001 - 0.002, $p$ = 3 - 5). The quantity $S(f,p) \equiv  \bigg( 1 - \frac{\chi^{2} - \chi^{2}_{min}}{\chi^{2}_{max} - \chi^{2}_{min}}\bigg)$ is in the range 0 - 1. The maximum value of $S(f, p)$ corresponds to the minimum $\chi^{2}$ value. In the plot we have shown $S(f,p)$ contours where $S(f,p)$ $\geq 0.97$ is considered as the allowed range (the red region) for the two parameters and from the plot we determine the value of $p$ = $3.81 \pm 0.004$ and $f$ = $(1.23 \pm 0.09)\times 10^{-3}$.


\begin{figure}[H]
\begin{center}
\includegraphics[scale=0.28]{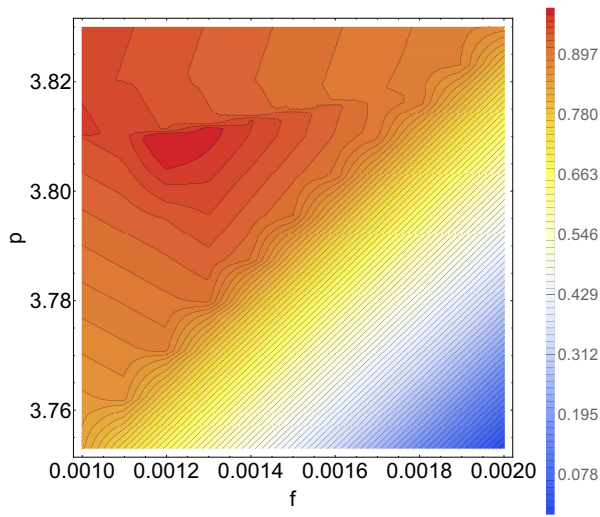}
\caption{The $S(f, p)$ plot for determination of $p$ and $f$ is shown where, the maxima (minima of $\chi^{2}$) occurs at $p$ = $3.81 \pm 0.004$ and $f$ = $(1.23 \pm 0.09)\times 10^{-3}$.\label{fig:chisq}}
\end{center}
\end{figure}

\section{Summary of results and discussion \label{s3}}
We summarize and discuss our key results below.
\begin{enumerate}
\item Assuming the $M_{\bullet} - \sigma$ relation, $M_{\bullet} = f M_{b}$ and  a single power law profile for the stellar mass density we have analytically shown that $p(\gamma) = (2\gamma + 6)/(2 + \gamma)$ (eq. (\ref{th_pl_f})). For a typical a range of $\gamma$ = 0.75 - 1.4, we find $p$ = 3.6 - 5.3, which is within the observed range.
\item The second analysis shown is for the Nuker profile which is a double power law with two slopes $\beta$ and $\Gamma$. As an approximate analysis, we take an average value of the mean slope to be $(\beta + \Gamma)/2$ for the set of 12 galaxies (tabulated in Table \ref{data_nuker}) resulting  $p= 3.86$ from eq. (\ref{pl_slope}), where the value of $p$ obtained from $\chi^{2}$ analysis is 3.81, which is very close. Therefore, our analysis of observational data agrees well with the theoretical expected value.
\item We have described a procedure for determining the $M_{\bullet} - \sigma$ relation (see Fig. \ref{flowchart}). Previous models (as discussed in the \S \ref{s1}) determined the $M_{\bullet} - \sigma$ relation and also the $M_{b} - M_{\bullet}$ relation independently. We have determined those two relations self - consistently in our model from our $\chi^{2}$ analysis (see eq. (\ref{chisq})).
\item For power law galaxies, we started directly from mass density profile, which in the case of Nuker profile was obtained by inverting the intensity profiles. The $\sigma$ for different power law indices are shown in Fig. \ref{fig:plsigma}. The variation of $p$ and $\log k$ with different values of $\gamma$, $r_{s}$ and $M_{s}$ are shown in Fig. \ref{fig:plslint_a} and Fig. \ref{fig:plslint_b}. 
The variation of $\sigma$ with different $M_{\bullet}$ and $M_{s}$ is shown in Fig. \ref{pl_cont2_a} and Fig. \ref{pl_cont2_b}. For a fixed value of $\gamma$, $\log k$ and $\sigma$ depend on the value of $M_{s}$ and $r_{s}$. These various diagnostics enable us to interpret the relation by using the observables such as $\gamma$ and $M_{s}$ and to predict $k$ and $p$. From observational Nuker intensity profiles, we have determined the LOS velocity dispersion of the stars in the galaxy through their distribution function (see Fig.  \ref{fig:ngc_3115_a}, \ref{fig:ngc_3115_b}). By using a proportionality relation between $M_{b}$ and $M_{\bullet}$ we have derived values of $p$ and $\log k$ for different values of $f$ by a linear fit (see Fig. \ref{fig:m_sigma_fit_b} for scatter plot and Fig. \ref{fig:m_sigma_fit_a} for the linear fit for a fixed $f$) and through $\chi^{2}$ minimization (see Fig. \ref{fig:chisq}) for the Nuker case. From the scatter plot (see Fig. \ref{fig:m_sigma_fit_b}) it is seen even for a small set of galaxies that the $p$ and $\log k$ values within a specific range of $f$ are very close to the observed range; these are consistent with observations. The obtained values are $p$ = $3.81 \pm 0.004$ and $f$ = $(1.23 \pm 0.09)\times 10^{-3}$.
\end{enumerate}

\section{Conclusions \label{s4}}
We have discussed a procedure of deriving the $M_{\bullet} - \sigma$ relation along with a proportionality relation of $M_{b}$ and $M_{\bullet}$ starting from observational data by deriving the distribution function $f(\epsilon)$. Using our novel approach we can also determine the index of the nonlinear relation between $M_{b}$ and $M_{\bullet}$ (as mentioned earlier) as well as $p$ self consistently. The $M_{\bullet} - \sigma$ relation is complicated to explain by existing models. The self consistent determination of $f$, $k$ and $p$ is key for improving the models. The resolution of the problem can come from a DF $f(\epsilon, L_{z})$ built for a central BH and constraining a self - consistent dynamical model from which an explanation of $f M_{b} = k \sigma^p$ can finally emerge. That needs much more sophisticated analytical and numerical methods applied to both the bulge mass scaling as well as $M_{\bullet} - \sigma$ determination.

\bibliography{references}
\end{document}